\documentclass[aps,prb,preprint,groupedaddress,showpacs,floatfix]{revtex4-1}

\usepackage{amsmath,amssymb,latexsym,epic,eepic,epsfig,graphicx,psfrag}
\usepackage[latin1]{inputenc}
\begin{document}

\title{Electron transport  across a  metal-organic interface}

\author{Kurt Stokbro}
\email{kurt.stokbro@quantumwise.com}
\author{S\o ren Smidstrup}
\email{soren.smidstrup@quantumwise.com}
\affiliation{QuantumWise A/S,\\
 Lers{\o} Parkall\'{e} 107,
  DK-2100 Copenhagen, Denmark}
\homepage{http://quantumwise.com}

\date{\today}

% max 75 words
\begin{abstract}
We simulate the electron transport across the  Au(111)-pentacene
interface using non-equilibrium Green's functions and density-functional
theory (NEGF-DFT), and calculate the bias-dependent
electron transmission.  We find that the electrical contact resistance  is dominated by the
formation of a Schottky barrier at the interface, and show that the
 conventional semiconductor transport
models across  Schottky barriers need to be modified in order to
describe the simulation data.  We present an extension
of  the conventional Schottky barrier transport model, which can
 describe our simulation results and
 rationalize recent experimental data.
\end{abstract}

%suggested keywords

\maketitle
\section{Introduction}
Organic materials for electronics is a rapidly growing area, with
new commercial organic devices for applications in lighting, displays and photovoltaics being considered. An important
problem is the control of charge injection at the metal-organic
interface. The charge injection can be associated with a large contact
resistance, and is sometimes more dominating for the electrical
performance than the transport within the organic semiconductor. The
transport across the interface is usually described by the theory of
metal-semiconductor contacts\cite{Ishii-1999,
  Braun-2009, Kumatani-2013, Liu-2010}, where the transport is characterized by an
injection barrier that needs to be overcome by thermionic emission\cite{Sze-1981}. In this paper we
show that this theory cannot explain first-principles data of quantum transport  across a
metal-organic semiconductor interface. Organic semiconductor crystals have
much narrower electron bands than "traditional" inorganic semiconductors, and the theory must be extended to take
into account that for certain electron injection energies there may be no available organic crystal bands in
the band bending region,  and the electron needs to tunnel through this region.

Previous theoretical studies of the metal-organic interface have focused on understanding the
properties of a single or few layers of organic molecules on metal
surfaces. The focus in this
paper is to simulate a true interface between a  gold and a pentacene
bulk crystal through the use of density-functional theory (DFT) and
the non-equilibrium Green's function (NEGF) method.  To
our knowledge this is the first study of a single metal-organic interface
which does not make use of a slab geometry, but models semi-infinite
electrodes by  applying  open boundary conditions. We  show how
such simulations can give new insight into the electrostatic
properties of the interface, the contact resistance, and electron
transport across the interface.

We have
chosen the gold-pentacene crystal interface as our model system. Due to its high hole mobility, the
pentacene crystal  is an important organic
electronic material,  and the
gold-pentacene interface is one of the most well studied systems both
theoretically\cite{Ortega2011,Li-2009,Lee-2007,Lee-2005,Toyoda-2010,Saranya-2012}
and experimentally
\cite{Koch-2007,McDonald-2006,Schroeder-2002,Ihm-2006,France-2003,Kafer-2007,
  Watkins-2002,Kang-2003,Soe-2009,Diao-2007,Liu-2010, Park-2005},
thus there is a large number of
experimental and theoretical data for verification of the theoretical
simulations.

The organisation of the paper is the following. In
section~\ref{sec:method} we describe the computational model used for
the calculations, and in section~\ref{sec:results} we present the results
of the calculations. Section~\ref{sec:discussion} presents a simple
model which rationalize the results and the results are summarized in section~\ref{sec:summary}.

\section{Methodology}
\label{sec:method}
For all calculations we have used
Atomistix ToolKit (ATK)~\cite{ATK12.8}, which is a density-functional
theory code that uses numerical localized atom centered basis sets and
normconserving pseudopotentials. For the exchange-correlation potential we have used the generalized
gradient approximation (GGA) of Wang and Perdew\cite{pw91} (PW91) as
suggested by Li {\em et al.}\cite{Li-2009}

The electronic structure is expanded in basis sets optimized
to reproduce hydrogen and carbon dimer total energies following the procedure of  Blum
{\em et al.}\cite{Blum-2009} For carbon we  use 21 orbitals per atom with
$s$, $p$ and $d$ character and radial ranges up to 3.9~\AA, and for hydrogen we use
5 orbitals per atom with $s$ and $p$ character  and ranges up to 4.2~\AA.
For gold we use a minimal basis set with 9 orbitals per atom and
ranges up to 3.6~\AA, and add a layer of gold ghost orbitals above the
gold surface.  With this model we calculate an ionization energy
of 6.34~eV (6.589\cite{Gruhn-2002}) for pentacene, and a work function of 5.20~eV (5.26\cite{Hansson-1978}) for the Au(111)
surface, where the corresponding experimental values are given in parentheses.

To describe  the gold-pentacene interface we correct for basis set superposition
errors (BSSE) using the counterpoise correction\cite{Lee-2007}, and using this computational model with an ($8\times3$) k-point
grid,  we reproduce the geometry, work
function change, and adsorption energy\cite{Stokbro-2013} of pentacene on the
Au(111)-($\sqrt{3}\times6)$ surface obtained with a plane-wave
method\cite{Li-2009}. In this paper we study  pentacene on the
Au(111)-($2\times3\sqrt{3})$ surface. This structure has been
observed experimentally\cite{Koch-2007} and the adsorption energy of a pentacene
monolayer is higher for this structure compared with the
Au(111)-($\sqrt{3}\times6)$ surface\cite{Stokbro-2013}.

\section{Results}
\label{sec:results}
\begin{figure}[tbp]
\begin{center}
  \includegraphics[width=\linewidth]{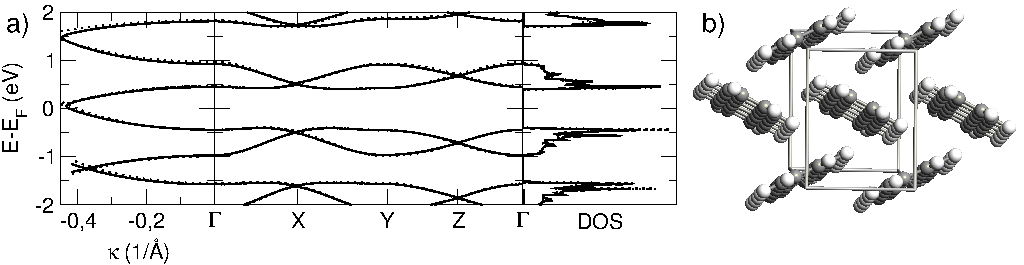}
\end{center}
  \caption{ The solid lines show the complex bandstructure, band
    structure and DOS of the relaxed pentacene crystal. The dotted
    lines show the corresponding values for the strained crystal which
    is used for the right electrode of the Au(111)-pentacene  interface
    configuration. }
  \label{fig:crystal}
\end{figure}

The experimentally observed  pentacene crystal geometry\cite{Sciefer-2006} is shown in
Fig.~\ref{fig:crystal}b. To combine the crystal with the Au(111)-($2\times3\sqrt{3})$ cell, it is necessary to
strain the pentacene crystal $\sim3$\% in the interface plane.
With this constraint on the cell, we then optimize the length of the interface cell in the transport
direction and the pentacene coordinates.
To investigate the influence of such modifications on the electronic
structure we have calculated the
complex band structure, band structure and density of states of the
pentacene crystal at the experimental lattice constant and compared
with the  strained crystal. The results
are shown in Fig.~\ref{fig:crystal}a, where the dotted lines are for
the strained case. We see that the electronic structure is almost identical to
the unstrained pentacene crystal (solid lines), and thus the straining of the crystal
will only have a minor impact on our electron transport calculations.

\begin{figure}[tbp]
\begin{center}
  \includegraphics[width=\linewidth]{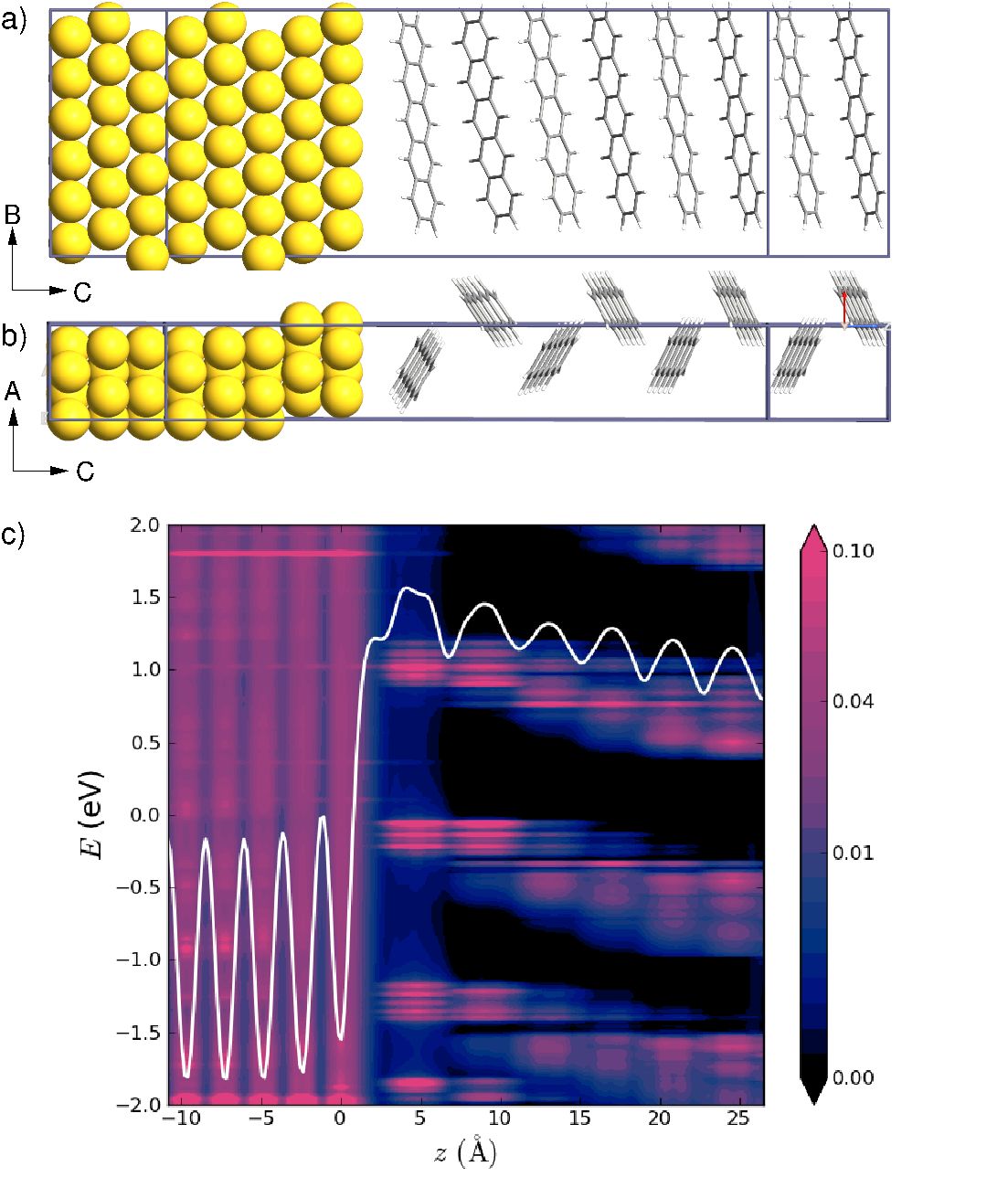}
\end{center}
  \caption{In our notation, the Au(111)-pentacene interface is spanned by
    the crystal cell vectors A$||x$ and B$||y$, while the transport direction is C$||z$.
    Side views of the a) BC and the b) AC planes of the
    interface are shown here. The
    outermost parts show the geometry of the semi-infinite
    electrodes. c)  Contour plot  of the local density of states (LDOS) of the
    Au-pentacene interface and the electrostatic
    potential (white line) along the $z$ direction. The energy $E$ is
    given relative to
    the electrode Fermi levels (which coincide at zero bias), and the LDOS and the potential
    are averaged over the $xy$-plane. An $(8\times3)$ k-point grid
    was used for the LDOS calculation.}
  \label{fig:device8}
\end{figure}

To set up the geometry of the  Au(111)-pentacene interface, we have
first relaxed an  Au(111)-$(2 \times 3 \sqrt{3})$ slab with 2
pentacene layers, until the forces of all
pentacene atoms and the first 2 gold layers were
below 0.01 eV/\AA. We next  used the pentacene
crystal as a template for extending the slab  from
2 to 6  pentacene layers, and relaxed all atoms in layers 2,
3, and 4 until the forces were below 0.02 eV/\AA. Finally, we attached semi-infinite
electrodes to set up a device configuration as shown in
Fig.~\ref{fig:device8}.  The BSSE correction cannot be applied in the device configuration,
and we can therefore only obtain reliable forces for
atoms in pentacene layers 2, 3, and 4.  In these layers the forces are
below 0.04 eV/\AA, confirming that our procedure for generating the
geometry using a slab configuration is accurate. In the following we will calculate the
properties of this interface.

Figure~\ref{fig:device8}c shows the local density of states (LDOS) of
the Au(111)-pentacene interface plotted along the transport direction $z$.
The 5 gold layers and 6 pentacene layers are visible as $z$ positions
with large LDOS values. For the pentacene layers we clearly see the molecular
levels. We note the shifts in the molecular levels in the $z$
direction, corresponding to a band bending in the organic crystal at the interface. The
solid white line shows the average electrostatic potential, which exhibits the
same band bending behaviour.

Figure~\ref{fig:device8}c also shows how the  metallic gold states extend
through the adsorbed pentacene molecules into the pentacene
crystal. This is seen as an asymmetric  shape of the LDOS states, with
a long tail at energies above the maximum in the LDOS peak. This is
particularly clear for molecules in the 3rd layer above the interface. Thus, the LDOS
illustrates the difficulties for the metallic states to propagate
through the band bending region into the pentacene electrode.

The band bending is a result of Fermi level pinning of the pentacene
highest occupied molecular orbital (HOMO) at the
gold-pentacene interface, which shifts the pentacene levels at the
interface relative to their positions in the organic crystal electrode.
The electronic structure in this (the right) electrode corresponds to an intrinsic
pentacene  bulk crystal where the Fermi level is in the middle of the band
gap. When no bias is applied between the two electrodes, the gold and pentacene  Fermi levels are
aligned, and the pentacene HOMO band edge in the right electrode is $\sim 0.4$~eV
below the gold Fermi level. At the interface the HOMO is pinned to
the gold Fermi level, thus creating a total band bending of 0.4 eV.

The pinning of the HOMO level arises because the ionization energy of
the pentacene crystal is lower than the work function of the gold
surface. We calculate a work function of gold of 5.20~eV and an
ionization energy of the pentacene crystal of 5.04~eV.

\begin{figure}[tbp]
\begin{center}
  \includegraphics[width=\linewidth]{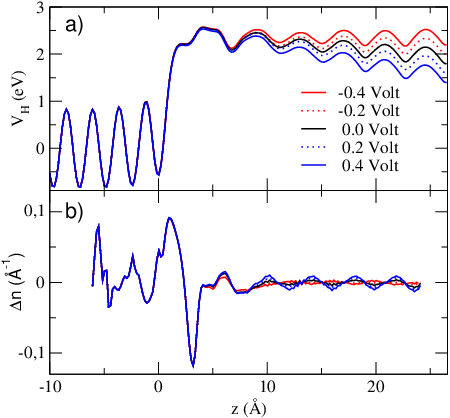}
\end{center}
  \caption{a) Electrostatic potential of the gold-pentacene
    interface averaged over the $xy$-plane for right electrode
    voltages $U_R$ =-0.4,
    -0.2, 0.0, 0.2, and 0.4~Volt. b) Change in electron density
    upon adsorption of pentacene on gold, integrated over the $xy$-plane, for the same
    bias voltages.}
  \label{fig:potbias}
\end{figure}

Figure~\ref{fig:potbias}b shows the change in the electron density of the
gold interface upon adsorption of pentacene. The plot was obtained by
subtracting the electron density of the isolated gold and pentacene
surfaces from the combined system, and shows the formation of  a
surface dipole at the interface. Note that this is not a charge transfer from pentacene to
gold, since the dipole is located between the top gold layer and the
first pentacene layer. The surface dipole arises from the so-called
pillow effect\cite{Koch-2007b}, where the pentacene molecule pushes the gold
density back.

This surface dipole lowers the workfunction of gold. For a single layer
of pentacene on Au(111) we calculate a work
function of 4.48~eV, which  is in excellent agreement with experimental
data  (4.52\cite{Schroeder-2002} , 4.4\cite{Watkins-2002} ,
4.6\cite{Diao-2007} eV).

We next apply different bias voltages, $U_R$ = -0.4,  -0.2,  0.0,
 0.2, and 0.4~Volt to the pentacene electrode. The
 bias will shift the electrochemical potential in the organic
 electrode by $\mu_R =
\mu_L-e U_R$ relative to the chemical potential $\mu_L$ in the metal
electrode. Figure~\ref{fig:potbias}a shows  the
electrostatic potential along the $z$-direction for the different bias voltages. The zero-bias
potential corresponds to the electrostatic potential of
Fig.~\ref{fig:device8}c. A negative bias increases the pentacene
electrochemical potential $\mu_R$ and thereby reduces the band bending, while a positive bias lowers $\mu_R$
which increases the band bending.

\begin{figure}[tbp]
\begin{center}
  \includegraphics[width=\linewidth]{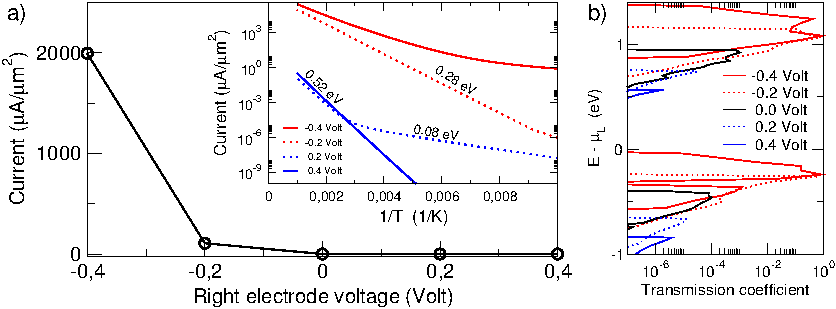}
\end{center}
  \caption{a) Current-voltage characteristics of the gold-pentacene
     interface. The inset shows the temperature dependence of the
    current at each bias voltage. b) Transmission spectra for each bias voltage.
    The transmission spectra are averaged
    over an $(8\times 3)$ k-point grid, and the the energy scale is relative to the left
    electrochemical potential $\mu_L$.}
  \label{fig:iv}
\end{figure}

Figure~\ref{fig:iv}a shows the calculated current-voltage characteristics
of the interface. The interface shows a strong rectifying behaviour.
The inset in the figure also shows the temperature dependence
of the current. The temperature dependence shows activated electron
transport, meaning that thermionic emission is  the dominating current
contribution. For each bias the barrier corresponding to the slope of the curve
is indicated. In the following we will investigate this in
further detail and make a simple model of the transport across the
interface.

Figure~\ref{fig:iv}b shows the transmission spectrum as function of
the bias voltage applied to the right electrode. The transmission increases
exponentially when a negative voltage is applied. At $-0.4$~V the
maximum transmission is 1, corresponding to a very good coupling
between the pentacene crystal and the gold electrode. It follows from the
discussion of the band bending above that the
system is in flat band condition at this bias voltage, which can also be seen in
Fig.~\ref{fig:potbias}. For the other computed bias points the bands bend
downwards, and a Schottky-type barrier arises at the interface.

In an inorganic semiconductor, a Schottky barrier is reflected by an energy
shift of the transmission peaks relative to the metal Fermi level;
however, the magnitude of the transmission peaks will not change.
For the pentacene
crystal, the transmission peaks are both shifted and reduced in magnitude. The
reduction is an effect of the band bending of the narrow molecular
bands. The band bending is of similar magnitude as the widths of the
organic bands, which has the effect that  the electron
cannot propagate inside a molecular band  all the way from the
organic crystal to the metal electrode. Outside a molecular band the
wavefunction amplitude is exponentially damped with the tunneling
distance, thus, for increasing band bending
the transmission peak is exponentially lowered.

\begin{figure}[tbp]
\begin{center}
  \includegraphics[width=\linewidth]{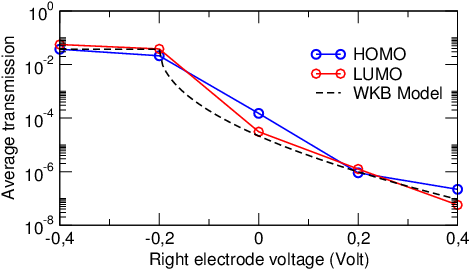}
\end{center}
  \caption{The average transmission of the HOMO (blue line)
    and the LUMO (red line) transmission peaks. A simple WKB model of
    the transmission is shown as the dashed line.}
  \label{fig:homolumo-current}
\end{figure}

\section{Discussion}
\label{sec:discussion}
Finally, we present a model that quantifies the reduction of the transmission peaks in the
inorganic crystal due to the tunneling through the band bending
region. Figure~\ref{fig:homolumo-current} shows
the transmission averaged over the HOMO and the LUMO peaks. The curve
is flat up to $-0.2$~V, and then decreases exponentially. The dashed
black line shows a WKB model of the transmission, where the average
transmission is
given by
\begin{equation}
T \propto e^{-2 d \sqrt{\frac{2 m}{\hbar^2}\bar{\phi}}}.
\label{eq:T}
\end{equation}
In this equation, $d = 23$~\AA\ is the distance from the first
pentacene layer to the right electrode, and $\bar{\phi} =
  \frac{1}{2} e (U_R - U_0)$ is an effective tunnel barrier,
  corresponding to vanishing barrier at $U_0=-0.2$~V and a linearly
  increasing barrier
  $e (U_R - U_0) z/d$ as a function of $z$.

The total current at bias $eU_R=\mu_L-\mu_R$ is given by
\begin{equation}
I \propto  \int T(E) \left[ f(\frac{E-\mu_L}{k_B T}) - f(\frac{E-\mu_R}{k_B T})  \right] dE,
\end{equation}
where $T(E)$ is the transmission coefficient, $f$ the Fermi function,
$T$ the electron temperature, and $k_B$ the Boltzmann constant.

Under forward bias the current can be approximated by
\begin{equation}
I \propto T(E_v) \, e^{-\phi/k_B T},
\end{equation}
where $E_v=\mu_L-\phi$ and  $\phi$ is the offset between the semiconductor valence band and
the metal chemical potential. For an inorganic semiconductor
$T(E_v)\sim 1$, but for an organic semiconductor
$T(E_v)\sim e^{-\alpha d}$ as inferred from Eq.~(\ref{eq:T}).

\begin{figure}[tbp]
\begin{center}
  \includegraphics[width=\linewidth]{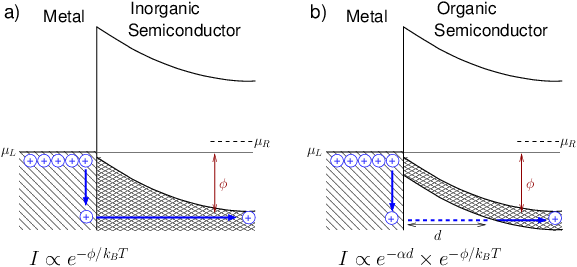}
\end{center}
  \caption{Band diagram of a Schottky barrier under forward bias for a) metal-inorganic semiconductor
    and b) metal-organic semiconductor interface. Due to the finite width of the
  organic crystal bands, the electron transmission is exponentially
  damped when propagating over the Schottky barrier of the
  metal-organic interface. }
  \label{fig:banddiagram}
\end{figure}

This qualitative difference in the electron transmission across a metal-semiconductor interface for
inorganic and organic systems, respectively, is illustrated in
Fig.~\ref{fig:banddiagram}. The model illustrates how the narrow organic
crystal bands leads to an additional  Schottky barrier
contact resistance for a metal-organic interface.

It is interesting to compare this model  with the experimental data
for transport across the gold-pentacene interface by Liu {\it et al.}\cite{Liu-2010} They  find a
large contact resistance and rectification for the gold-pentacene
interface,  in qualitative agreement with our calculations. (A
quantitative comparison is not immediately possible since the effective
contact area is unknown in the experiment.) Liu {\it et al.}\ model the data in
terms of a traditional Schottky barrier. In order to explain the large
contact resistance they suggest that the interface dipole rigidly
shifts the bands in the organic semiconductor, which introduces an
additional injection barrier. We note, however, that their model
assumes a shift of  the Fermi level in the organic crystal
relative to its  HOMO and LUMO bands upon forming the interface with
gold, thus, the interface dipole should  change the organic crystal
from p to n type, which we find unphysical.

Liu {\it et al.}\ also assume band
bending at the interface, however, in their case the band bending has
no direct electrical effect on the contact resistance. Our
model suggests that the band bending will indeed affect the injection
probability, thus leading to an alternative model for the large
contact resistance which does not require the introduction of an
additional injection barrier.

Furthermore, Liu {\it et al.}\ noticed that the metal-organic contact resistance can
be reduced by incorporating an inorganic semiconductor buffer layer between the
metal and the organic crystal\cite{Kumatani-2013, Liu-2010}. They
suggest that the effect of the inorganic semiconductor is to remove
the interface dipole and thereby the additional injection barrier.
We propose an alternative explanation for this effect. The buffer
layer will unpin the HOMO band at the interface, making the
organic crystal bands more flat, which increases the electron
propagation probability in the band bending region. Thus, unlike  models
proposed previously\cite{ Kumatani-2013, Liu-2010}, we suggest that
the buffer layers will not change the Schottky barrier,
but rather increases the injection probability due to flatter
inorganic bands which provides more efficient
propagation through the band bending region.

\section{Summary}
\label{sec:summary}
In summary, we have studied bias-induced electron transport across the Au(111)-pentacene
interface. The study shows that there is a good chemical contact between the
gold and the pentacene crystals, and the electrical resistance is dominated
by  a Schottky  barrier within the organic
crystal. The transport across the Schottky barrier is thermally
activated, however transport is reduced compared to an
inorganic semiconductor, since the electron needs to tunnel through
part of the band bending region, as illustrated in
Fig.~\ref{fig:banddiagram}. Our calculation is a simplification compared to
an experimental situation; in particular, the length of the band
bending region is orders of magnitude shorter than it is
experimentally. However, the calculations suggest an overlooked
effect, namely that propagation through the band bending region can be
damped due to the narrow bands in the organic semiconductor. We have
illustrated how such a  model can rationalize recent
experimental data for the contact resistance of the gold-pentacene
interface and explain the effect of an inorganic buffer layer on the contact resistance.

\begin{acknowledgments}\label{sec:acknowledgements}
We acknowledge Alexander Bratkovski for discussions on the
metal-organic interface and Anders Blom for proof-reading the manuscript.
\end{acknowledgments}

\bibliography{paper}

\end{document}